\documentclass[english]{article}
\usepackage[T1]{fontenc}
\usepackage[latin9]{inputenc}
\usepackage{geometry}
\usepackage{array}
\usepackage{float}
\usepackage{multirow}
\usepackage{amssymb}
\usepackage{graphicx}

\makeatletter

\providecommand{\tabularnewline}{\\}

\makeatother

\usepackage{babel}
\begin{document}

\title{Combining dependent p-values resulting from multiple effect size
homogeneity tests in meta-analysis for binary outcomes}

\author{Osama Almalik\thanks{Researcher, Department of Mathematics and Computer Science, Eindhoven
University of Technology. E-mail: o.almalik@tue.nl.}}
\maketitle
\begin{abstract}
\noindent Testing effect size homogeneity is an essential part when
conducting a meta-analysis. Comparative studies of effect size homogeneity
tests in case of binary outcomes are found in the literature, but
no test has come out as an absolute winner. A alternative approach
would be to carry out multiple effect size homogeneity tests on the
same meta-analysis and combine the resulting dependent p-values. In
this article we applied the correlated Lancaster method for dependent
statistical tests. To investigate the proposed approach's performance,
we applied eight different effect size homogeneity tests on a case
study and on simulated datasets, and combined the resulting p-values.
The proposed method has similar performance to that of tests based
on the score function in the presence of a effect size when the number
of studies is small, but outperforms these tests as the number of
studies increases. However, the method's performance is sensitive
to the correlation coefficient value assumed between dependent tests,
and only performs well when this value is high. More research is needed
to investigate the method's assumptions on correlation in case of
effect size homogeneity tests, and to study the method's performance
in meta-analysis of continuous outcomes.
\end{abstract}

\section{Introduction}

Meta-analysts still carry out effect size homogeneity tests on a regular
basis \cite{key-1-1-1-1-1-1,key-1-1-1-1-1-2}. Several methods have
been developed to test effect size homogeneity in meta analysis with
multiple 2x2 contingency tables, and the performance of these methods
have been studied in the literature. Jones et al. (1989) studied the
performances of the Likelihood Ratio test, Pearson's chi square test,
Breslow-Day test and Tarone's adjustment to it, a conditional score
test and Liang and Self's normal approximation to the score test \cite{key-1-1-1-1-1}.
Gavaghan et al. (1999) compared the performance of the Peto statistic,
the Woolf statistic, the Q-statistic (applied to the estimates of
the risk difference), Liang and Self's normal approximation to the
score test, and the Breslow-Day test \cite{key-1-1-1-1-2}. Almalik
and van den Heuvel (2018) compared the performance of the fixed effects
logistic regression analysis, the random effects logistic regression
analysis, the Q-statistic, the Bliss statistic, the $I^{2}$, the
Breslow-Day test, the Zelen statistic, Liang and Self's $T_{2}$ and
$T_{R}$ statistics, and the Peto statistic \cite{key-1-1-1-5-2}.
A recent study focused on meta-analysis with rare binary events \cite{key-1-1-1-1-4-1}.
However, no one specific test could be presented as a universal winner
from these comparative studies. An alternative approach would be to
perform multiple tests of effect size homogeneity on the same meta-analysis,
and combine the resulting p-values.

\noindent The earliest method to combine p-values resulting from independent
statistical tests is Fisher's method \cite{key-1-1-1-1}. Since then
many methods have been proposed to combine p-values resulting from
independent statistical tests, see \cite{key-1-1-1-2,key-1-1-1-2-1}
for an overview. The p-values in our context result from different
statistical tests for effect size homogeneity testing the same null
hypothesis. Since these p-values result from the same meta-analysis
dataset, these p-values as correlated \cite{key-1-1-1-6-1}. Therefore,
we only consider methods that combine p-values resulting from dependent
tests, and we present a brief review of these methods here.

\noindent Brown (1975) \cite{key-1-1-1-3} extended Fisher\textquoteright s
method to the case where the p-values result from test statistics
having a multivariate normal distribution with a known covariance
matrix. Kost and McDermott (2002) \cite{key-1-1-1-4} extended Brown\textquoteright s
method analytically for unknown covariance matrices. Other methods
to calculate the covariance matrix have been proposed in the literature
\cite{key-1-1-1-5}. Brown's method and its improvements can only
be applied to the case where the test statistics follow a multivariate
normal distribution. Since most test statistics used to test effect
size homogeneity approximately follow the Chi Square distribution,
these methods can not be applied.

\noindent Makambi (2003) modified the Fisher statistic, using weights
derived from the data, to accommodate correlation between the p-values
\cite{key-1-1-1-6-1}. However, the author applied methods developed
by Brown to derive the first two moments of the weighted distribution
and to estimate the correlation coefficients, which implies that the
same distributional assumptions made by Brown must hold here. Similar
modifications of the Fisher statistic can be found in the literature
\cite{key-1-1-1-6-2}. Yang (2010) introduced an approximation of
the null distribution of the Fisher statistic, based on the Lindeberg
Central Limit theorem. However, one condition that the test statistics
being m-dependent clearly does not hold here \cite{key-1-1-1-5-1}.
Another approximation introduced in Yang (2010) based on permutations
is said by the author to be numerically intensive \cite{key-1-1-1-5-1}.
Methods combining dependent p-values for very specific applications
have been proposed in the literature \cite{key-1-1-1-7-1,key-1-1-1-7-2,key-1-1-1-7-4},
and Bootstrap methods have been applied to this problem as well \cite{key-1-1-1-8-1}.

\noindent One method developed by Hartung (1999) considered dependent
test statistics testing one null hypothesis, each having a continuous
distribution under the null hypothesis \cite{key-1-1-1-6}. Using
the probability integral transformation \cite{key-1-1-1-8-1-1}, the
dependent p-values are transformed into standardized z values using
the probit function. Then the author proposed a formula for combining
the z values into one z value using weights for each z value and a
correlation coefficient between each two z values. This combined z
value is then used to test the original null hypothesis. Hartung's
method is only applicable for one-sided hypothesis tests, which makes
it applicable for combing p-values resulting from effect size homogeneity
tests based on random effects having one-sided alternative hypothesis.
However, this approach is not promising since effect size homogeneity
tests using random effects have been shown to perform poorly \cite{key-1-1-1-5-2,key-1-1-1-1-4-1}.

\noindent Another possibility is a method developed by Dai et al.
\cite{key-1-1-1-8,key-1-1-1-2-1-1}. Dai's approach uses a combined
test statistic developed earlier by Lancaster \cite{key-1-1-1-7},
but adjusted to incorporate correlations between the p-values. Lancaster
presented a test statistic that transforms the independent p-values
into a Chi squared test statistic using the inverse cumulative distribution
function of the Gamma distribution. Dai et al. (2014) noted that after
introducing correlation between the p-values the Lancaster statistic
no longer follows the Chi square distribution, and provided five approaches
to approximate the distribution of the Lancaster test statistic under
correlation. The approximation is done using the observed test statistics
and their corresponding degrees of freedom. The basic approach presented
by the authors is the Satterthewaite approximation \cite{key-1-1-1-9},
and the other four approaches are based on the Satterthwaite approximation
as well. The authors recommended using the Satterthwaite approximation
as the standard procedure to adjust the Lancaster statistic. See section
2.2.2 for a detailed description of the Sattherthwaite approximation
of the correlated Lancaster test statistic. 

\noindent This article is structured as follows. Section 2 presents
a short description of the tests for effect size homogeneity using
two-sided alternative hypothesis and of the correlated Lancaster procedure.
Section 3 describes the case study and the simulation model used.
The results are presented in section 4 and the discussion is relegated
to section 5.

\section{Statistical methods based on effect size homogeneity tests using
two-sided alternative hypothesis}

We first introduce notation that will be used throughout this section.
For a binary clinical outcome, let $X_{1i}$ and $X_{0i}$ be the
number of successes in the treatment group and the control group in
study $i$ (i$=1,2,...,m$) out of $n_{1i}$ and $n_{0i}$ trails,
respectively. Methods testing effect size homogeneity assume that
$E\left(X_{1i}\right)=n_{1i}\cdot p_{1i}$, with $p_{1i}$ the proportion
of success for the treatment group in study $i$. The Odds ratio is
calculated by $\hat{OR}_{i}=X_{1i}\left(n_{0i}-X_{0i}\right)/\left[\left(n_{1i}-X_{1i}\right)X_{0i}\right]$
and its standard error is given by $\hat{se}_{i}^{2}=\frac{1}{X_{1i}}+\frac{1}{n_{1i}-X_{1i}}+\frac{1}{X_{0i}}+\frac{1}{n_{0i}-X_{0i}}$.
All methods testing effect size homogeneity in this section assume
that the proportion $p_{ji}$, $j=0,1$, satisfy the following form:
$logit\left(p_{ji}\right)=\alpha_{i}+\left(\beta+\gamma_{i}\right)\cdot t_{ji}$,
with $\alpha_{i}$ an intercept for study $i$, $\beta$ the (mean)
effect size, $t_{ji}$ a treatment indicator variable for study $i$
with value 1 when $j=1$ and 0 otherwise, and $\gamma_{i}$ a study
treatment interaction effect for study $i$. Then effect size homogeneity
tests apply the following hypotheses

\noindent 
\begin{equation}
H_{0}:\gamma_{1}=\gamma_{2}=\ldots=\gamma_{m}=0\,\,\,vs\,\,H_{1}:\gamma_{i}\neq\gamma_{i^{'}}\:for\,some\,i\neq i^{'}.\label{eq:null_fixed}
\end{equation}

\subsection{Tests of effect size homogeneity }

\noindent In this section the effect size homogeneity tests based
on fixed-effects are briefly described. Under the null hypothesis
in (\ref{eq:null_fixed}), all these tests have an approximately Chi
squared distributed test statistic with $m-1$ degrees of freedom
\cite{key-1-1-1-8-2,key-1-1-1-8-3,key-1-1-1-8-4,key-1-1-1-8-5,key-1-1-1-8-6,key-1-1-1-8-7,key-1-1-1-8-8,key-1-1-1-8-9}.

\subsubsection{The Likelihood ratio test }

Assuming a fixed-effects logistic regression model, and using the
notation presented above, the success probability $p_{ji}$ for a
subject in study $i$ receiving treatment $j$, $j=0,1$, is given
by\linebreak{}
 $p_{ji}=\exp\left(\alpha_{i}+\left(\beta+\lambda_{i}\right)t_{ji}\right)/\left(1+\exp\left(\alpha_{i}+\left(\beta+\lambda_{i}\right)t_{ji}\right)\right)$.
The log Likelihood function can be constructed as follows

\begin{equation}
\begin{array}{lr}
l_{F}\left(\alpha_{i},\beta,\lambda_{i}\right)=\sum_{i=1}^{m}\left(X_{1i}\left(\alpha_{i}+\beta+\gamma_{i}\right)-n_{1i}\log\left(1+\exp\left(\alpha_{i}+\beta+\gamma_{i}\right)\right)\right)\\
\,\,\,\,\,\,\,\,\,\,\,\,\,\,\,\,\,\,\,\,\,\,\,\,\,\,\quad\,+\sum_{i=1}^{m}\left(X_{0i}\alpha_{i}-n_{0i}\log\left(1+\exp\left(\alpha_{i}\right)\right)\right).
\end{array}\label{eq:Fixed logistic}
\end{equation}

\noindent The Maximum Likelihood estimates are obtained under the
Full and the Null models, denoted from now on by indexes F and N,
respectively. The Likelihood ratio test statistic is given by \\
$T_{1}=-2\left(l_{N}\left(\hat{\alpha}_{i\left(N\right)},\hat{\beta}_{\left(N\right)},0\right)-l_{F}\left(\hat{\alpha}_{i\left(F\right)},\hat{\beta}_{\left(F\right)},\hat{\gamma}_{i}\right)\right)$.

\subsubsection{Tests based on the Q-statistic }

\noindent Defining $\hat{\beta}_{i}=\log\left(\hat{OR}_{i}\right)$
as the primary effect size, the Q-statistic \cite{key-1-1-1-8-2}
is given by $T_{2}=\sum_{i=1}^{m}\left(\hat{\beta}_{i}-\bar{\beta}\right)^{2}/\hat{se}_{i}^{2}$,
with $\bar{\beta}$ a weighted average given by $\bar{\beta}=\sum_{i=1}^{m}\left(\hat{\beta_{i}}/\hat{se}_{i}^{2}\right)/\sum_{i=1}^{m}\left(1/\hat{se}_{i}^{2}\right)$.
Bliss's test statistic is given by 
\[
T_{3}=\left(m-1\right)+\sqrt{\left(\bar{n}-4)\right)/\left(\bar{n}-1\right)}\left\{ \left(\bar{n}-2\right)\cdot T_{2}/\bar{n}-\left(m-1\right)\right\} 
\]
 with $\bar{n}=\left(\sum_{i=1}^{m}\left(n_{1i}+n_{0i}-2\right)\right)/m$
\cite{key-1-1-1-8-3,key-1-1-1-5-2}.

\subsubsection{Tests based on the score function}

\noindent The Breslow-Day approach \cite{key-1-1-1-8-4} adjusted
by Tarone \cite{key-1-1-1-8-6} can be described as follows. Firstly,
the Cochran-Mantel-Haenszel pooled odds ratio $OR_{C}$ is calculated
using \linebreak{}
 $\hat{OR}_{C}=\sum_{i=1}^{m}\left(X_{1i}\left(n_{0i}-X_{0i}\right)/n_{i}\right)/\sum_{i=1}^{m}\left(X_{0i}\left(n_{1i}-X_{1i}\right)/n_{i}\right)$.
Define $E_{C}\left(X_{1i}\right)=E\left(X_{1i}|X_{i},\hat{OR}_{C}\right)$
as the expected value of $X_{1i}$ given $X_{i}=X_{0i}+X_{1i}$, where
$E_{C}\left(X_{1i}\right)$ is obtained by solving the following equation

\begin{equation}
\left(\hat{OR}_{C}-1\right)E_{C}\left(X_{1i}\right)^{2}-\left(\left(X_{i}+n_{1i}\right)\hat{OR}_{C}+\left(n_{0i}-X_{i}\right)\right)E_{C}\left(X_{1i}\right)+X_{i}n_{1i}\hat{OR}_{C}=0\label{eq:Exp value}
\end{equation}

\noindent The Breslow-Day approach adjusted by Tarone statistic is
given by

\noindent 
\[
T_{4}=\sum_{i=1}^{m}\frac{\left(X_{1i}-E_{C}\left(X_{1i}\right)\right)^{2}}{var\left(X_{1i}|X_{i},\hat{OR}_{C}\right)}-\frac{\left(\sum_{i=1}^{m}X_{1i}-\sum_{i=1}^{m}E_{C}\left(X_{1i}\right)\right)^{2}}{\sum_{i=1}^{m}var\left(X_{1i}|X_{i},\hat{OR}_{C}\right)}.
\]

\noindent with $var\left(X_{1i}|X_{i},OR_{C}\right)$ is given by
\begin{equation}
var\left(X_{1i}|X_{i},\hat{OR}_{C}\right)=\left(\frac{1}{E_{C}\left(X_{1i}\right)}+\frac{1}{X_{i}-E_{C}\left(X_{1i}\right)}+\frac{1}{n_{1i}-E_{C}\left(X_{1i}\right)}+\frac{1}{n_{0i}-X_{i}+E_{C}\left(X_{1i}\right)}\right)^{-1}.\label{eq:variance}
\end{equation}

\noindent Zelen's test statistic \cite{key-1-1-1-8-6}, later corrected
by Halperin et al. \cite{key-1-1-1-8-7}, can be described as follows.
Firstly, the odds ratio $\hat{OR}_{Z}$ is given by $\hat{OR}_{Z}=\exp\left(\hat{\beta}_{\left(N\right)}\right)$
with $\hat{\beta}_{\left(N\right)}$ the Maximum Likelihood estimator
of the Fixed effects logistic regression analysis (\ref{eq:Fixed logistic})
under the null hypothesis of effect size homogeneity. The corrected
Zelen statistic for testing effect size homogeneity is given by \linebreak{}
$T_{5}=\sum_{i=1}^{m}\left(X_{1i}-E_{Z}\left(X_{1i}\right)\right)^{2}/var\left(X_{1i}|X_{i},\hat{OR}_{Z}\right)$,
with $E_{Z}\left(X_{1i}\right)$ now obtained by solving equation
(\ref{eq:Exp value}) with $\hat{OR}_{C}$ replaced by $\hat{OR}_{Z}$,
and $var\left(X_{1i}|X_{i},\hat{OR}_{Z}\right)$ is obtained by  equation
(\ref{eq:variance}) with $E_{C}\left(X_{1i}\right)$ replaced by
$E_{Z}\left(X_{1i}\right)$.

\noindent Liang and Self (1985) \cite{key-1-1-1-8-8} developed the
following test statistic using $\hat{OR}_{CL}=\exp\left(\hat{\beta}_{\left(CL\right)}\right)$,
with $\hat{\beta}_{\left(CL\right)}$ the Maximum Likelihood estimator
of the conditional Likelihood function given $X_{i}$ and $\gamma_{i}=0$,
$\forall i$. The the test statistic is given by $T_{6}=\sum_{i=1}^{m}\left(X_{1i}-E_{CL}\left(X_{1i}\right)\right)^{2}/var\left(X_{1i}|X_{i},\hat{OR}_{CL}\right)$,
with $E_{CL}\left(X_{1i}\right)$ now obtained by solving equation
(\ref{eq:Exp value}) with $\hat{OR}_{C}$ replaced by $\hat{OR}_{CL}$,
and $var\left(X_{1i}|X_{i},\hat{OR}_{CL}\right)$ is obtained by equation
(\ref{eq:variance}) with $E_{C}\left(X_{1i}\right)$ replaced by
$E_{CL}\left(X_{1i}\right)$. 

\subsubsection{Woolf statistic}

The Woolf statistic \cite{key-1-1-1-8-8-1} is given by $T_{7}=\sum_{i=1}^{m}\left(\hat{\beta_{i}}/\hat{se}_{i}\right)^{2}-\left(\sum_{i=1}^{m}\left(\hat{\beta_{i}}/\hat{se}_{i}^{2}\right)\right)^{2}/\sum_{i=1}^{m}\left(1/\hat{se}_{i}^{2}\right)$.

\subsubsection{Peto test}

\noindent The Peto statistic is given by \cite{key-1-1-1-8-9}

\noindent 
\[
T_{8}=\sum_{i=1}^{m}\frac{\left(X_{1i}-X_{i}n_{1i}/n_{i}\right)^{2}}{V_{i}}-\frac{\left(\sum_{i=1}^{m}\left(X_{1i}-X_{i}n_{1i}/n_{i}\right)\right)^{2}}{\sum_{i=1}^{m}V_{i}}
\]

\noindent where $n_{i}=n_{1i}+n_{0i}$ and $V_{i}=\left(X_{i}\left(n_{i}-X_{i}\right)n_{1i}n_{0i}\right)/\left(n_{i}^{2}\left(n_{i}-1\right)\right)$. 

\subsection{Correlated Lancaster procedure for combining correlated p-values}

In this section we describe the correlated Lancaster procedure \cite{key-1-1-1-8}
for combining dependent p-values resulting from the above mentioned
eight effect size homogeneity tests. Lancaster's method assumes there
are $n$ statistical tests each resulting in a test statistic $T_{i}$,
$i=1,\cdots.n$, degrees of freedom, $df_{i}$, and a p value, $p_{i}$.
Applying the probability integral transformation \cite{key-1-1-1-8-1-1},
it is noted that $1-p_{i}$ are uniformly distributed on $\left(0,1\right)$.
Lancaster showed that for $n$ independent p-values that $T=\sum\limits _{i=1}^{n}\gamma_{\left(df_{i}/2,2\right)}^{-1}\left(1-p_{i}\right)\sim\chi_{df}^{2}$
where $\gamma_{\left(df_{i}/2,2\right)}^{-1}$ is the inverse cumulative
distribution function of a Gamma distribution with a shape parameter
$df_{i}/2$ and a scale parameter 2, and $df=\sum_{i=1}^{n}df_{i}$
\cite{key-1-1-1-7}. Dai et al. (2014), noting that for correlated
p-values the $T$ statistic does not follow a $\chi_{df}^{2}$ anymore,
suggested five methods to approximate the distribution of $T$. The
authors recommended a method using the Satterthwaite approximation
\cite{key-1-1-1-9} as a standard procedure to adjust the $T$ statistic,
which can be described as follows. For a set of correlated p-values,
the authors noted that $E\left(T\right)=\sum\limits _{i=1}^{n}df_{i}=df$,
and $Var\left(T\right)=2\sum\limits _{i=1}^{n}df_{i}+2\sum\limits _{i<k}\mathop{c}ov_{ik}$
with $cov_{ik}=cov\left(\gamma_{\left(df_{i}/2,2\right)}^{-1}\left(1-p_{i}\right),\gamma_{\left(df_{k}/2,2\right)}^{-1}\left(1-p_{k}\right)\right)$.
Next the authors defined the statistic $T_{A}=cT\thickapprox\chi_{\nu}^{2}$
where $c=\nu/E\left(T\right)$ and $\nu=2\left[E\left(T\right)\right]^{2}/Var\left(T\right)$,
where $c$ and $\nu$ are chosen so that the first and second moments
of the scaled chi-square distribution and the distribution of $T$
under the null are identical \cite{key-1-1-1-2-1,key-1-1-1-2-1-1}.
The statistic $T_{A}$ can be used for testing the null hypothesis
of effect size homogeneity. The authors presented several methods
to estimate $\rho_{ik}$ \cite{key-1-1-1-8}.

\section{Case study and simulation model}

\subsection{Case study}

Bein et al. (2021) carried out a systematic review and meta-analysis
to investigate the risk of adverse pregnancy, perinatal and early
childhood outcomes among women with subclinical hypothyroidism treated
with Levothyroxince \cite{key-1-1-1-8-14}. Among the extensive study
was a meta-analysis of preterm delivery associated with levothyroxine
treatment versus no treatment among women with subclinical hypothyroidism
during pregnancy. This meta-analysis included seven studies, each
study having a group treated with Levothyroxine and a control group.
For each group the number of preterm delivery (events) was noted.
This meta-analysis was used here as a case study and the data are
shown in Table 1. 

\begin{table}[H]
\caption{Meta-analysis on studying the effect of Levothyroxine on preterm pregnancy
among women with subclinical hypothyroidism (Bein et al. 2021) }

\begin{centering}
\begin{tabular}{|l|c|c|c|c|}
\hline 
\multirow{2}{*}{Study} & \multicolumn{2}{c|}{Levothyroxine} & \multicolumn{2}{c|}{Control}\tabularnewline
\cline{2-5} 
 & Events & Total & Events & Total\tabularnewline
\hline 
\hline 
Casey et al. (2017) & 40 & 339 & 47 & 338\tabularnewline
\hline 
Maraka et al. (2016) & 4 & 82 & 30 & 284\tabularnewline
\hline 
Maraka et al. (2017) & 60 & 843 & 236 & 4562\tabularnewline
\hline 
Nazarpour et al. (2017) & 4 & 56 & 14 & 58\tabularnewline
\hline 
Nazarpour et al. (2018) & 18 & 183 & 21 & 183\tabularnewline
\hline 
Wang et al. (2012) & 0 & 28 & 9 & 168\tabularnewline
\hline 
Zhao et al. (2018) & 7 & 62 & 6 & 31\tabularnewline
\hline 
\end{tabular}
\par\end{centering}
\end{table}

\subsection{Simulation model}

The simulation model applied can be described as follows \cite{key-1-1-1-8-13,key-1-1-1-5-2}.
In total $m$ studies are created, and for the $i^{th}$ study $n_{ji}$,
$j=0,1$, are independently drawn from a Poisson distribution with
parameter $\delta$. The random variables $X_{ji}$ are drawn independently
from the Binomial distribution $Bin\left(n_{ji},p_{ji}\right)$, with
$p_{0i}=\exp\left(\alpha_{i}\right)/\left(1+\exp\left(\alpha_{i}\right)\right)$
and $p_{1i}=\exp\left(\alpha_{i}+\beta+\gamma_{i}\right)/\left(1+\exp\left(\alpha_{i}+\beta+\gamma_{i}\right)\right)$.
Here $\alpha_{i}$ is study-specific intercept, $\beta$ is a constant
effect size, and $\gamma_{i}$ is study-specific effect size with
$\alpha_{i}\sim\left(\alpha,\sigma_{\alpha}^{2}\right)$ and $\gamma_{i}\sim N\left(0,\tau^{2}\right)$.
Two scenario's are considered: effect size homogeneity ($\tau^{2}=0$
) and effect size heterogeneity ($\tau^{2}>0$). The following parameter
values were used in the simulation study: $m=5,10,20,30$, $\delta=50$,
$\alpha=0$, $\beta=0,2$, $\sigma_{\alpha}^{2}=1$, $\tau^{2}=0,0.15,0.3,0.5$.
A number of 1000 simulation runs was carried out for each parameter
combination, and the average Type I error rate and the average statistical
power based on significance level of 0.05 were calculated and presented
in the results section.

\section{Results}

This section presents the results of the case-study and the simulation
study. For the correlated Lancaster method we used $\rho_{ik}=0.25,0.5,0.75$. 

\subsection{Case study}

\noindent We applied the Breslow-Day test adjusted by Tarone (BDT),
the Bliss test, the Liang \& Self test, the Likelihood ratio test
(LRT), the Peto test, the Q-statistic (Q), the Woolf test and the
Zelen test to test the effect size homogeneity hypothesis for the
meta-analysis study in Bein et al. (2021). Subsequently, we used the
correlated Lancaster method (CORR. LANC.) to combine the eight resulting
p-values. The resulting p-values are shown in Table 2. All homogeneity
tests and the correlated Lancaster method (for all values of $\rho_{ik}$)
rejected the effect size homogeneity hypothesis (p<0.05). The effect
size homogeneity tests based on the score function and the Peto test
produced similar p-values. The Q-statistic and the Bliss statistic
produced substantially lower p-values than all other tests. For the
correlated Lancaster method the p-value increased as the $\rho_{ik}$
value increased.
\begin{table}[H]
\caption{p-values from the effect size homogeneity tests and the correlated
Lancaster method applied to meta-analysis from Bein et al. (2021)}

\centering{}%
\begin{tabular}{|>{\raggedright}m{1cm}|>{\raggedright}m{1.6cm}|>{\raggedright}m{2cm}|}
\hline 
\multicolumn{2}{|>{\raggedright}p{0.52cm}|}{Method} & p-value\tabularnewline
\hline 
\multicolumn{2}{|>{\raggedright}p{1.5cm}|}{BDT} & 0.007514 \tabularnewline
\hline 
\multicolumn{2}{|l|}{BLISS} & 0.000003283\tabularnewline
\hline 
\multicolumn{2}{|l|}{LIANG \& SELF} & 0.007486 \tabularnewline
\hline 
\multicolumn{2}{|l|}{LRT} & 0.004327 \tabularnewline
\hline 
\multicolumn{2}{|l|}{PETO} & 0.008340 \tabularnewline
\hline 
\multicolumn{2}{|>{\raggedright}p{1.5cm}|}{Q} & 0.000003122 \tabularnewline
\hline 
\multicolumn{2}{|l|}{WOOLF} & 0.020232 \tabularnewline
\hline 
\multicolumn{2}{|l|}{ZELEN} & 0.007485 \tabularnewline
\hline 
\multirow{3}{1cm}{CORR. LANC.} & $\rho_{ik}=0.25$ & 0.000000353 \tabularnewline
\cline{2-3} 
 & $\rho_{ik}=0.5$ & 0.000043090\tabularnewline
\cline{2-3} 
 & $\rho_{ik}=0.75$ & 0.000371 \tabularnewline
\hline 
\end{tabular}
\end{table}

\subsection{Results of simulation study}

Table 3 shows the average Type 1 error rates of the Breslow-Day test,
the Bliss test, the Liang \& Self test, the Peto test, the Q-statistic,
the Woolf test, the Zelen test and the correlated Lancaster method.
For the correlated Lancaster method, it is noted that the value of
the correlation coefficient between the p-values resulting from effect
size homogeneity test statistics affects the Type I error values,
with the Type I error closest to the nominal value when $\rho_{ik}=0.75$.
In case of no effect size, all homogeneity tests are conservative,
while the combined Lancaster method is liberal. This pattern is consistent
for the different number of studies included in a meta-analysis. In
case $\beta=2$, all effect size homogeneity tests with the exception
of the Liang \& Self, Likelihood ratio test and the Zelen tests, have
a Type I error value below the nominal value. The Likelihood ratio
test has a Type I error rate above the nominal level for all values
of $m$. The Liang \& Self has a Type I error rate equaling the nominal
value when $m=5,10$. The Zelen test has a nominal Type I error value
when $m=5$ but a Type I error value above the nominal value when
$m=10$. As $m$ increases, the Type I error rates of the Liang \&
Self and the Zelen tests increase way above the nominal value. The
Q-statistic and the Bliss statistic have a Type I error rates lower
than the nominal value, and these rates decrease as $m$ increases.
When $\rho_{ik}=0.75$ and for all values of $m$, the correlated
Lancaster method has a Type I error either equal to or below the nominal
value.

\begin{table}[H]
\caption{Type I error for fixed-effects homogeneity tests and the correlated
Lancaster method $\tau^{2}=0$}

\centering{}%
\begin{tabular}{|>{\raggedright}m{1cm}|>{\raggedright}m{1.6cm}|>{\raggedright}m{1cm}|>{\raggedright}m{1cm}|>{\raggedright}m{1cm}|>{\raggedright}m{1cm}|>{\raggedright}m{1cm}|>{\raggedright}m{1cm}|>{\raggedright}m{1cm}|>{\raggedright}m{1cm}|}
\hline 
\multicolumn{2}{|>{\raggedright}p{0.5cm}|}{} & \multicolumn{2}{c|}{$m=5$} & \multicolumn{2}{c|}{$m=10$} & \multicolumn{2}{c|}{ $m=20$} & \multicolumn{2}{c|}{ $m=30$}\tabularnewline
\hline 
\multicolumn{2}{|>{\raggedright}p{0.52cm}|}{} & $\beta=0$ & $\beta=2$ & $\beta=0$ & $\beta=2$ & $\beta=0$ & $\beta=2$ & $\beta=0$ & $\beta=2$\tabularnewline
\hline 
\multicolumn{2}{|>{\raggedright}p{1.5cm}|}{BDT} & 0.044 & 0.045 & 0.049 & 0.045 & 0.049 & 0.051 & 0.040 & 0.068\tabularnewline
\hline 
\multicolumn{2}{|l|}{BLISS} & 0.040 & 0.036 & 0.041 & 0.022 & 0.028 & 0.021 & 0.023 & 0.022\tabularnewline
\hline 
\multicolumn{2}{|l|}{LIANG \& SELF} & 0.044 & 0.049 & 0.049 & 0.050 & 0.049 & 0.054 & 0.040 & 0.070\tabularnewline
\hline 
\multicolumn{2}{|l|}{LRT} & 0.048 & 0.064 & 0.055 & 0.059 & 0.055 & 0.077 & 0.049 & 0.094\tabularnewline
\hline 
\multicolumn{2}{|l|}{PETO} & 0.041 & 0.013 & 0.045 & 0.004 & 0.048 & 0.002 & 0.037 & 0.002\tabularnewline
\hline 
\multicolumn{2}{|>{\raggedright}p{1.5cm}|}{Q} & 0.047 & 0.042 & 0.046 & 0.023 & 0.039 & 0.024 & 0.028 & 0.027\tabularnewline
\hline 
\multicolumn{2}{|l|}{WOOLF} & 0.037 & 0.032 & 0.041 & 0.016 & 0.037 & 0.016 & 0.028 & 0.017\tabularnewline
\hline 
\multicolumn{2}{|l|}{ZELEN} & 0.044 & 0.050 & 0.049 & 0.053 & 0.049 & 0.055 & 0.040 & 0.074\tabularnewline
\hline 
\multirow{3}{1cm}{CORR. LANC.} & $\rho_{ik}=0.25$ & 0.135 & 0.099 & 0.148 & 0.107 & 0.144 & 0.094 & 0.138 & 0.102\tabularnewline
\cline{2-10} 
 & $\rho_{ik}=0.5$ & 0.089 & 0.069 & 0.103 & 0.062 & 0.090 & 0.063 & 0.077 & 0.070\tabularnewline
\cline{2-10} 
 & $\rho_{ik}=0.75$ & 0.057 & 0.050 & 0.071 & 0.036 & 0.060 & 0.044 & 0.054 & 0.050\tabularnewline
\hline 
\end{tabular}
\end{table}

\noindent The statistical power is shown in figures 1, 2 and 3. Since
the Type I error was closest to the nominal value when $\rho_{ik}=0.75$,
the statistical power of the correlated Lancaster method is only shown
for the case $\rho_{ik}=0.75$. When there is no effect size, $\beta=0$,
and in the case of low heterogeneity, $\tau^{2}=0.15$, all methods
have a statistical power lower than 30\% when $m=5$, with the correlated
Lancaster method showing higher statistical power than other methods.
Of the other methods, the Likelihood ratio test, Liang and Self test
and Breslow-Day test have slightly higher statistical power, followed
by Zelen, Peto, Q-statistic, Woolf and Bliss. The same pattern remains
as $m$ increases, with the statistical power of all methods increasing
but remaining below than 80\%. In case of moderate and high heterogeneity
($\tau^{2}=0.3,0.5$), the statistical power increases for all methods
when $m=5$, and all methods approach or surpass the nominal value
of 95\% as $m$ increases. The correlated Lancaster method still has
higher statistical power than other methods.

\noindent When $\beta=2$, $\tau^{2}=0.15$ and $m=5$, the Likelihood
ratio test has the highest statistical power. The Likelihood ratio
test is followed by the Zelen, Liang and Self, BDT and the correlated
Lancaster methods. The Q-statistic, Bliss, Woolf and Peto methods
have clearly a lower statistical power. All methods, however, have
a statistical power below 25\%. As $m$ increases, the statistical
power of all methods increases slightly, except for the Peto method
which remains below 10\%. As the heterogeneity level increases, $\tau^{2}=0.3$,0.5,
the statistical power increases for all methods when $m=5$. The same
pattern persists with the Likelihood ratio, Zelen, Liang and Self,
Breslow-Day tests and the correlated Lancaster methods still having
the highest power, although the statistical power remains below 50\%.
As $m$ increases, the statistical power increases for all methods.
When $\tau^{2}=0.5$ the Likelihood ratio test's statistical power
exceeds the nominal level, while the Zelen, Liang and Self, Breslow-Day
and the correlated Lancaster methods have statistical power close
or equal to the nominal value when $m=30$. 

\noindent 
\begin{figure}[H]
\includegraphics[scale=0.4]{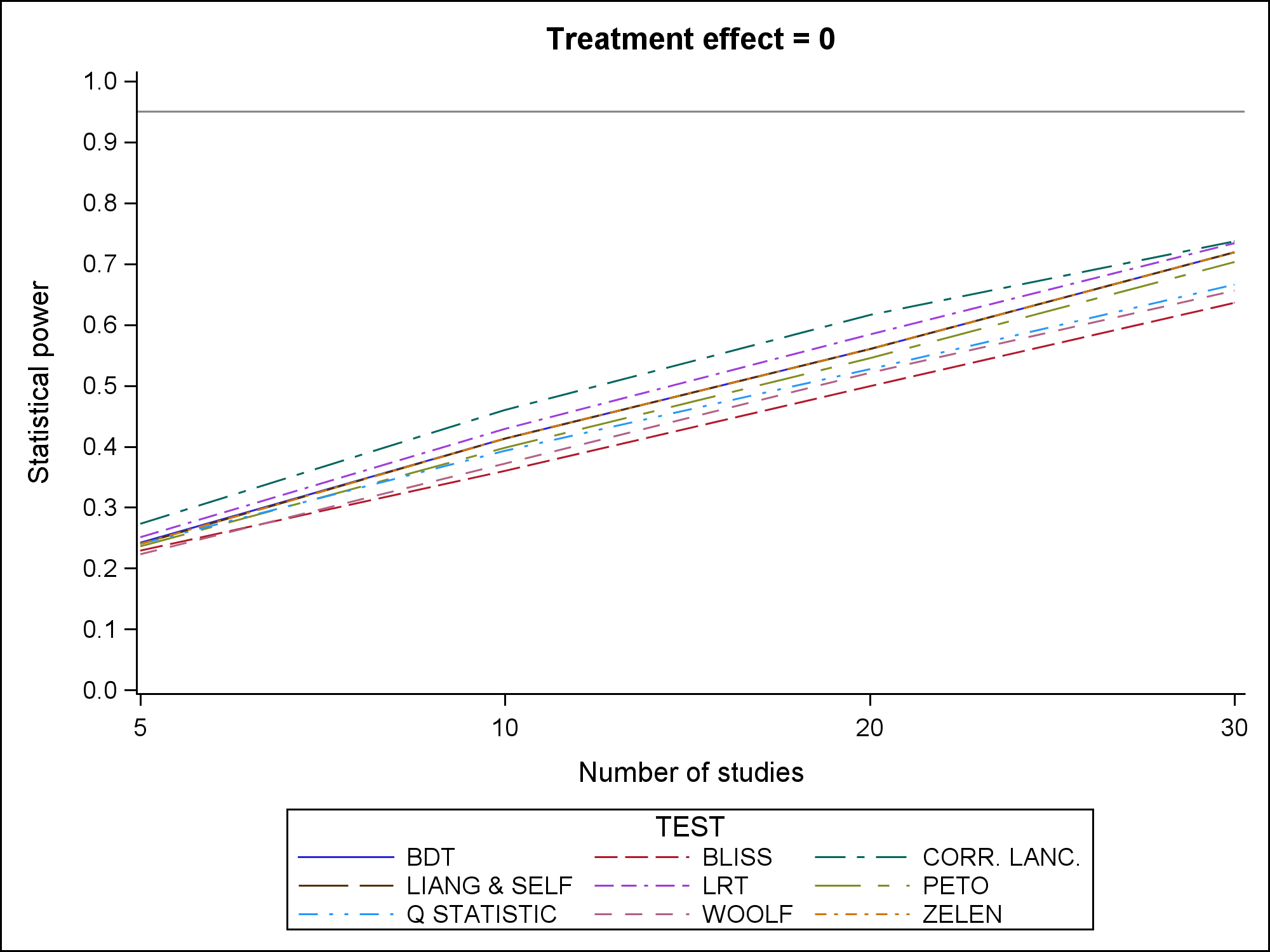}\includegraphics[scale=0.4]{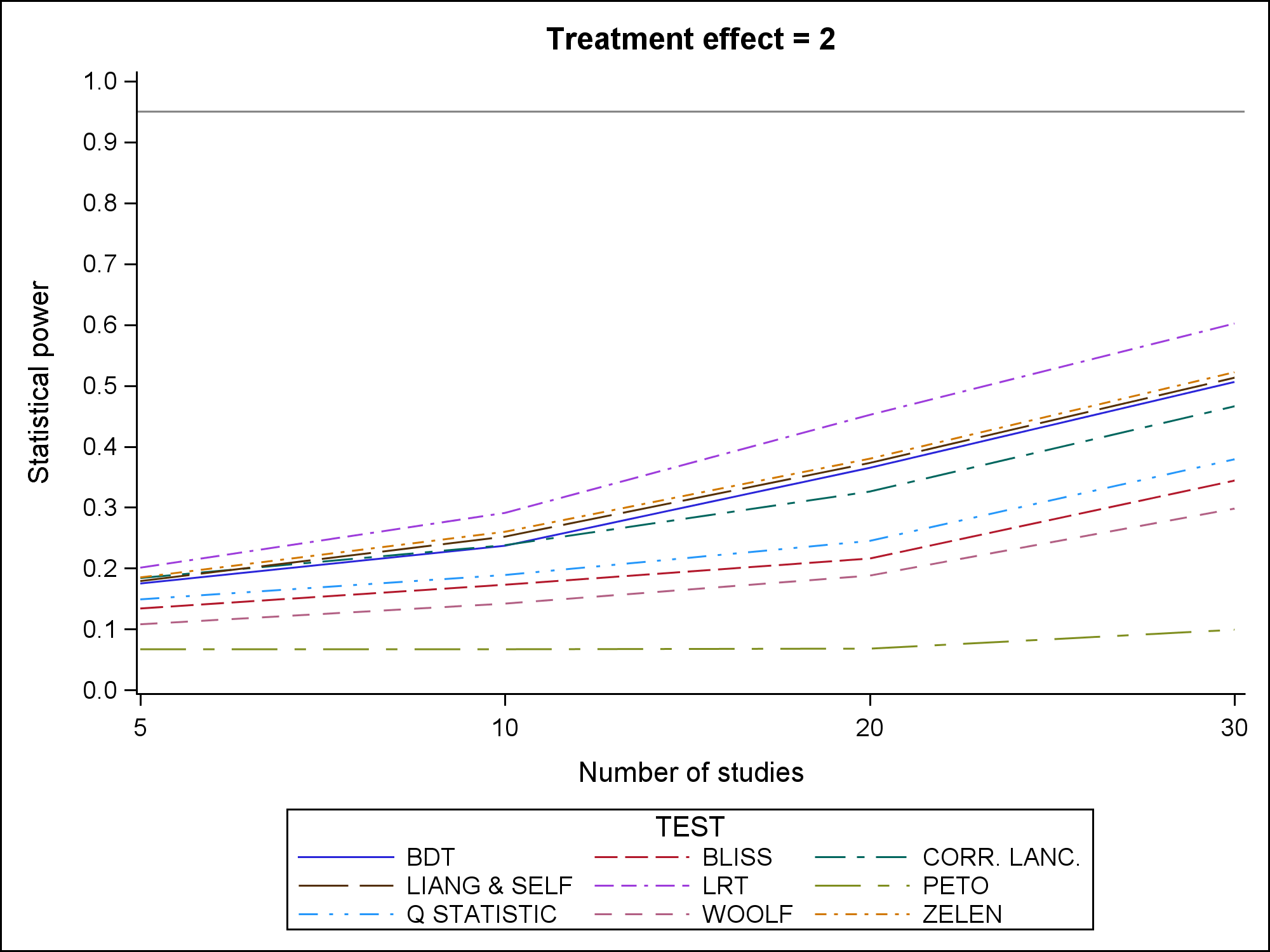}

\caption{Statistical power in case weak effect size heterogeneity: $\tau^{2}=0.15$}
\end{figure}
\begin{figure}[H]
\includegraphics[scale=0.4]{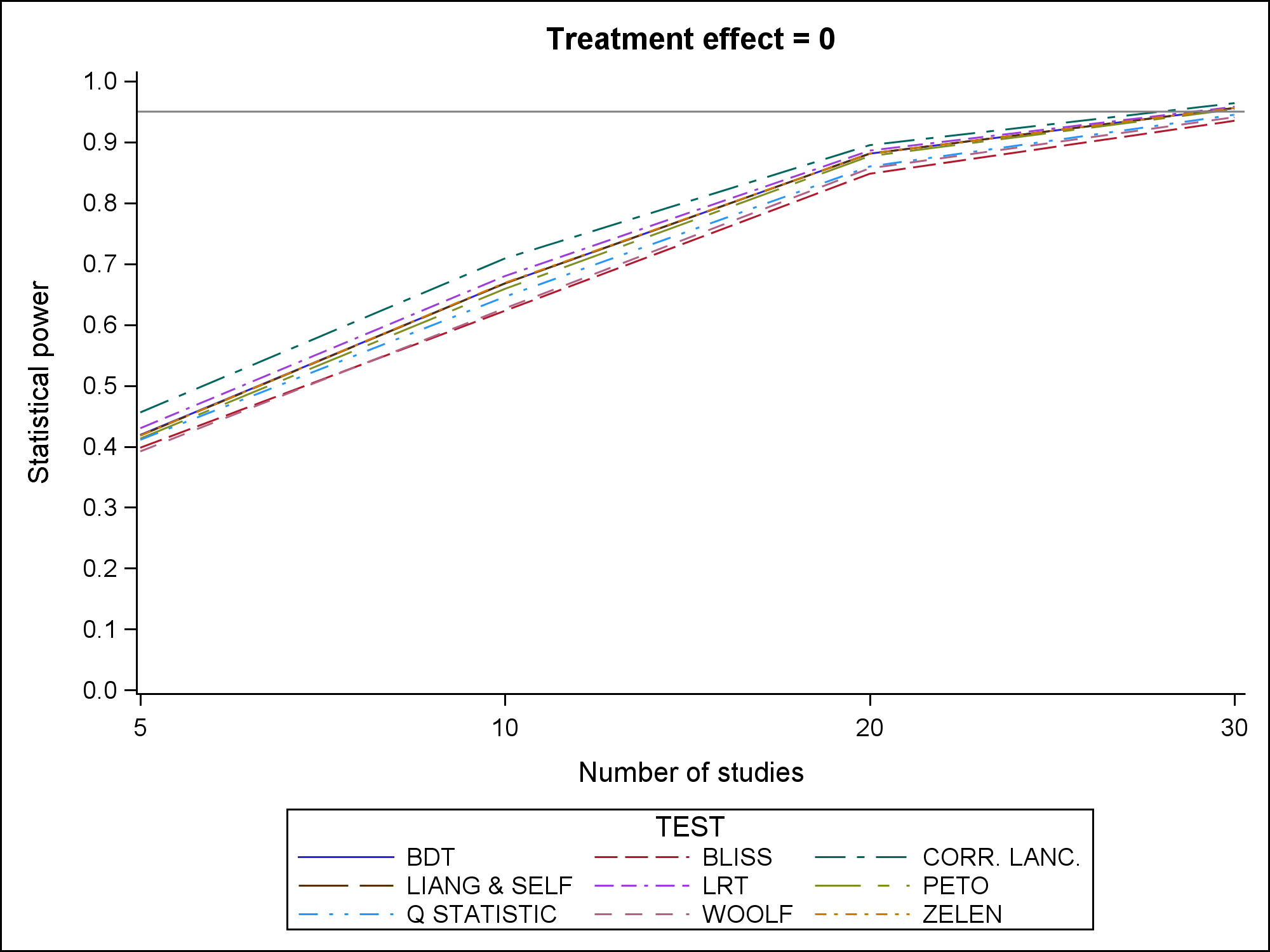}\includegraphics[scale=0.4]{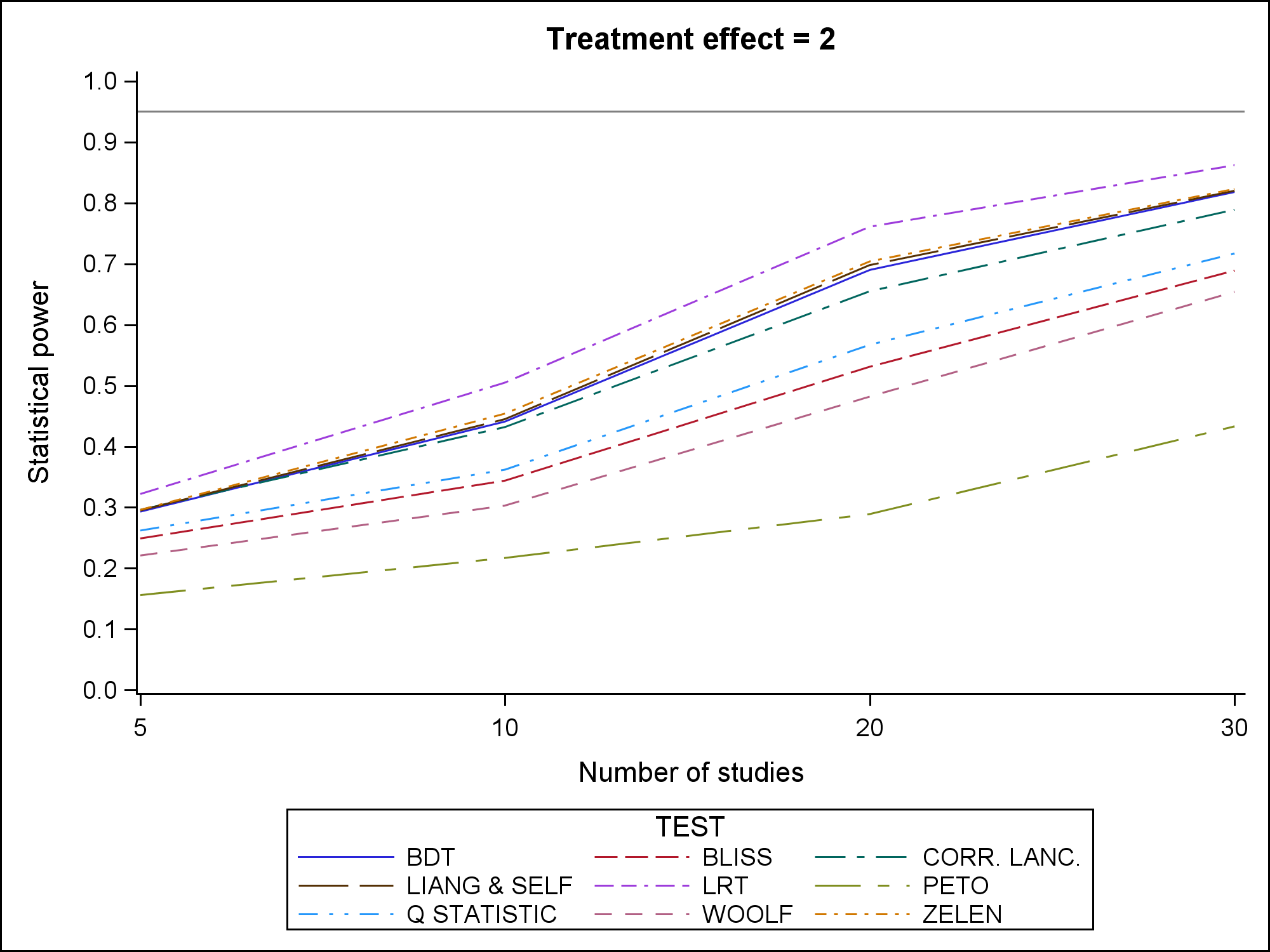}

\caption{Statistical power in case moderate effect size heterogeneity: $\tau^{2}=0.3$}
\end{figure}
\begin{figure}[H]
\includegraphics[scale=0.4]{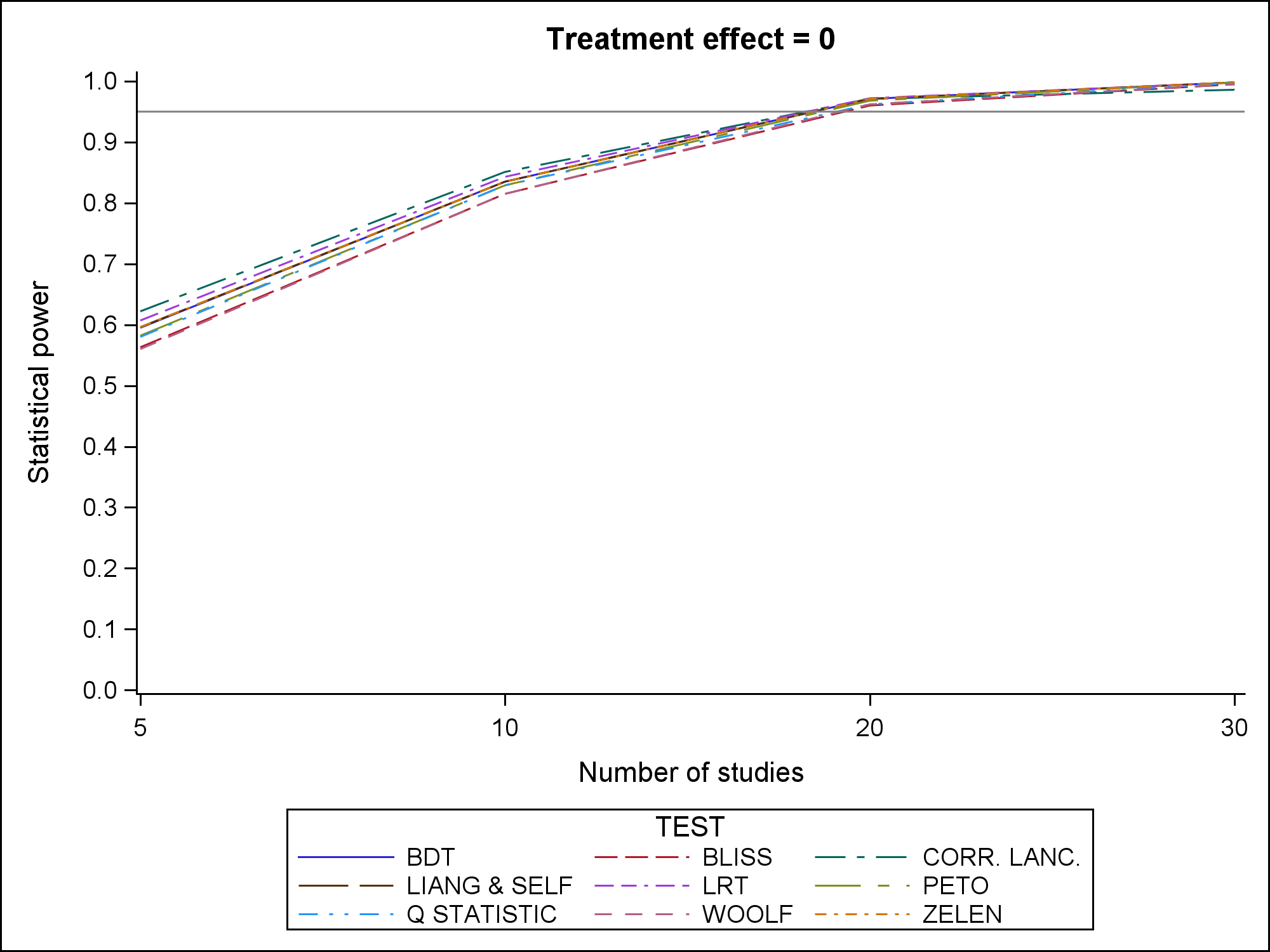}\includegraphics[scale=0.4]{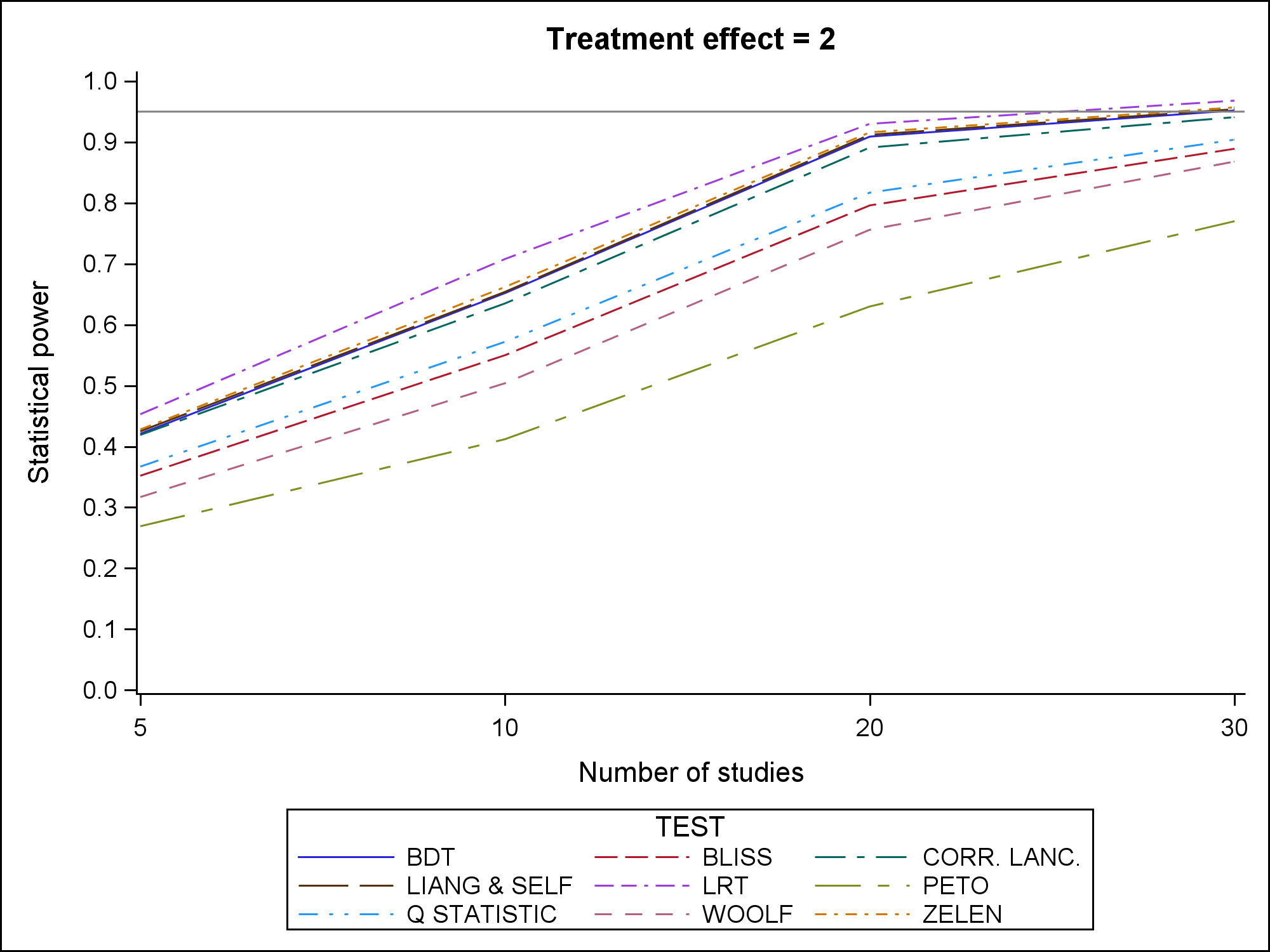}

\caption{Statistical power in case strong effect size heterogeneity: $\tau^{2}=0.5$}

\end{figure}

\section{Discussion}

The purpose of this article was applying an approach to combine p-values
of different effect size homogeneity tests applied to a meta-analysis
of binary outcomes. The proposed approach is an adjustment of a method
introduced in Lancaster (1961) to combine independent p-values into
a Chi squared test statistic. The Lancaster method was adjusted to
incorporate correlation between dependent p-values. The Satterthewaite
method was used to approximate the correlated Lancaster test statistic
in case of dependent p-values. The method was originally developed
for aggregating effects in high-dimensional genetic data analysis
(Dai et al. 2012). To study the performance of the proposed method
we analyzed a real life meta-analysis, and we carried out a simulation
study with multiple scenarios including different number of studies,
different effect sizes and different levels of heterogeneity. We tested
the null hypothesis of effect size homogeneity using the Breslow-Day
test, the Bliss test, the Liang \& Self test, the Likelihood ratio
test, the Peto test, the Q-statistic, the Woolf test, and the Zelen
test for the case study and for the simulated datasets. Subsequently,
we combined the resulting p-values from these eight effect size homogeneity
tests using the correlated Lancaster method. For the case study we
compared the performance of the correlated Lancaster method to that
of the eight effect size homogeneity tests using the p-values. For
the simulation study we did the comparison using the average Type
I error rates and the average statistical power. 

\noindent Some findings regarding the effect size homogeneity tests
have been established earlier in the literature. The Likelihood ratio
test is liberal, and tests based on the score function (Breslow-Day,
Liang \& Self and Zelen tests) perform well when the number of studies
is small \cite{key-1-1-1-1-1,key-1-1-1-5-2}. However, as the number
of studies increases these score function tests tend become liberal
\cite{key-1-1-1-5-2}. The Q-statistic and the Bliss statistic are
conservative and they get more conservative as the number of studies
increases \cite{key-1-1-1-1-1,key-1-1-1-5-2}. 

\noindent The correlated Lancaster method is sensitive to the value
of correlation coefficient between the test statistics, as the combined
p value is positively correlated with the value of the correlation
coefficient. This can be explained by the fact that higher values
of the correlation coefficient result in a larger variance of the
correlated Lancaster statistic. This produces a smaller value of the
correction constant $c$ and in turn a smaller value of the correlated
Lancaster test statistic and thereby a larger p value. The correlated
Lancaster method performs best in case of a high positive correlation
between the dependent test statistics, namely a correlation coefficient
value of 0.75. The correlated Lancaster method performs quite well
in the presence of a effect size, having a Type 1 error rate always
within the nominal value. Unlike all effect size homogeneity tests
considered here, the correlated Lancaster method is robust to the
number of studies in a meta-analysis. The statistical power of the
correlated Lancaster method is similar to that of the Breslow-Day,
Liang \& Self and the Zelen tests when the number of studies is small.
As the number of studies increases, these three tests have superior
statistical power to the correlated Lancaster method. This can be
explained by the inflated Type I error the Breslow-Day, Liang \& Self
and the Zelen tests when the number of studies increases. 

\noindent The correlated Lancaster method performs well and it is
easy to implement, but few reservations need to be mentioned. The
assumption of positive correlation between the dependent test statistics
is intuitively a reasonable assumption. However, the method's performance
is sensitive to the value of the correlation coefficient, as the method
performs best when the value of the correlation coefficient is high.
More research is needed to investigate the correlation levels between
dependent effect size homogeneity tests. In addition, we only applied
the method to balanced meta-analysis of binary outcomes. Extra research
is warranted to investigate the method's performance on unbalanced
meta-analysis, meta-analysis of continuous outcomes and meta-analysis
of rare binary events.

\part*{Conflict of interest}

The author has declared no conflict of interest.

\newpage{}

\section{Appendix: code used for implementing the correlated Lancaster method}

\noindent DATA DATASET\_CORR; 

\noindent SET P\_VALUES\_INI; 

\noindent DF = \&NUMBER\_OF\_CENTERS - 1; 

\noindent SHAPE = DF/2; SCALE = 2;

\noindent INV\_CDF = SQUANTILE('GAMMA',P\_VALUE,SHAPE,SCALE);

\noindent RUN;

\noindent PROC SORT DATA = DATASET\_CORR;

\noindent BY SIM TEST; 

\noindent RUN;

\noindent PROC MEANS DATA = DATASET\_CORR NOPRINT; 

\noindent VAR INV\_CDF DF; 

\noindent OUTPUT OUT = T\_VALUE SUM = T\_VALUE SUM\_DF; 

\noindent BY SIM;

\noindent RUN;

\noindent PROC MEANS DATA = P\_VALUES\_INI NOPRINT; 

\noindent VAR P\_VALUE; 

\noindent OUTPUT OUT = NUM\_TESTS N = NUM\_TESTS;

\noindent BY SIM; 

\noindent RUN;

\noindent DATA RHO; 

\noindent SET NUM\_TESTS(KEEP = SIM NUM\_TESTS);

\noindent COMB\_NR = NUM\_TESTS{*}(NUM\_TESTS - 1)/2; 

\noindent RUN;

\noindent DATA RHO\_S; 

\noindent SET RHO; 

\noindent DO COMB\_NR = 1 TO COMB\_NR; 

\noindent RHO = \&RHO{*}2{*}(\&NUMBER\_OF\_CENTERS - 1); 

\noindent OUTPUT; 

\noindent END; 

\noindent RUN;

\noindent PROC SORT DATA = RHO\_S; 

\noindent BY SIM;

\noindent RUN; 

\noindent PROC MEANS DATA = RHO\_S NOPRINT; 

\noindent VAR RHO; 

\noindent OUTPUT OUT = SUM\_RHOS SUM = SUM\_RHOS; 

\noindent BY SIM; 

\noindent RUN;

\noindent DATA LANC; 

\noindent SET T\_VALUE; 

\noindent P\_VALUE = 1 - PROBCHI(T\_VALUE,SUM\_DF); 

\noindent IF P\_VALUE <= \&SIG\_LEVEL THEN SIG = 1; 

\noindent ELSE SIG = 0; 

\noindent TEST = \textquotedbl LANCASTER\textquotedbl ; 

\noindent RUN;

\noindent DATA LANC\_CORR; 

\noindent MERGE T\_VALUE SUM\_RHOS; 

\noindent BY SIM; 

\noindent EXP\_T = SUM\_DF; 

\noindent VAR\_T = 2{*}SUM\_DF + 2{*}SUM\_RHOS; 

\noindent V = 2{*}(EXP\_T{*}{*}2)/VAR\_T; 

\noindent C = V/EXP\_T; 

\noindent T\_A = C{*}T\_VALUE; 

\noindent P\_VALUE = 1 - PROBCHI(T\_A,V); 

\noindent IF P\_VALUE <= \&SIG\_LEVEL THEN SIG = 1; 

\noindent ELSE SIG = 0; 

\noindent TEST = \textquotedbl LANC\_CORR\_POS\_CORR\textquotedbl ;

\noindent RUN;

\noindent DATA LANC\_TOTAL(KEEP = SIM TEST P\_VALUE SIG); 

\noindent SET LANC LANC\_CORR; 

\noindent BY SIM; 

\noindent RUN; 
\end{document}